\documentclass{entcs}
\usepackage{entcsmacro}
\usepackage{url}
\usepackage{graphicx}
\usepackage{moreverb}

\newcommand{\quid}{\hspace{0.5em}}
\def\lastname{Harcourt}
\begin{document}
\begin{frontmatter}
\title{Policies of System Level Pipeline Modeling}
 \author{Ed Harcourt\thanksref{myemail}}
  \address{Department of Mathematics, Computer Science, and Statistics \\
           St. Lawrence University\\
           Canton, NY USA}
    \thanks[myemail]{Email:
    \href{mailto:edharcourt@stlawu.edu} {\texttt{\normalshape
        edharcourt@stlawu.edu}}}
\begin{abstract}
%\footnotetext[1]{Department of Mathematics, Computer Science, and Statistics}
Pipelining is a well understood and often used implementation
technique for increasing the performance of a hardware system.
We develop several SystemC/C++
modeling techniques that allow us to quickly model, simulate, and
evaluate pipelines. We employ a small
domain specific language (DSL) based on resource usage patterns that
automates the drudgery of boilerplate code
needed to configure connectivity in simulation models. The
DSL is embedded directly in the host
modeling language SystemC/C++. Additionally we develop
several techniques for parameterizing a pipeline's behavior based
on {\em policies} of function, communication, and timing (performance modeling).
\end{abstract}
\end{frontmatter}

% patterns used: proxy, command, factory method, Strategy, interpreter

\section{\label{sec:intro}Introduction}
Pipelining is a well understood and often used implementation
technique for increasing the performance of hardware
\cite{kogge81,hwang84}. Since we have a taxonomy of pipeline designs
we can (and should) develop system-level
techniques that allow us to quickly model, simulate, and evaluate
various configurations.

In this paper we describe several modeling techniques inspired by
research in the generic and generative programming community
\cite{GPCE07,Czarnecki}. We use
SystemC \cite{systemc:lrm,glms:systemc} as our simulation framework
because of its support for system level modeling and simulation
and because it is
embedded in C++, a language with support for generic,
polymorphic, object oriented programming. Furthermore C++ is suitable
for constructing domain specific languages (DSLs) \cite{Abrahams05,Alexandrescu01}.

In system modeling simulation performance usually improves when we move to more abstract
models \cite{glms:systemc}. In software development it is often the opposite; abstract models suffer
an {\em abstraction penalty} \cite{veldhuizen99}. As a modeler
abstracts, performance may, at some point, begin to degrade. One goal is to ameliorate
the abstraction penalty by using compile time generic modeling techniques as opposed to
run-time techniques ({\em e.g.} virtual methods).

Pipelines are composed of {\em stages} that compute outputs at regular
intervals based on inputs. We'll start with a somewhat contrived example
of an application specific three stage linear pipeline that computes the
function $2x^2 + 4x - 7$.

\vspace{1em}
\mbox{
  \includegraphics[scale=0.5]{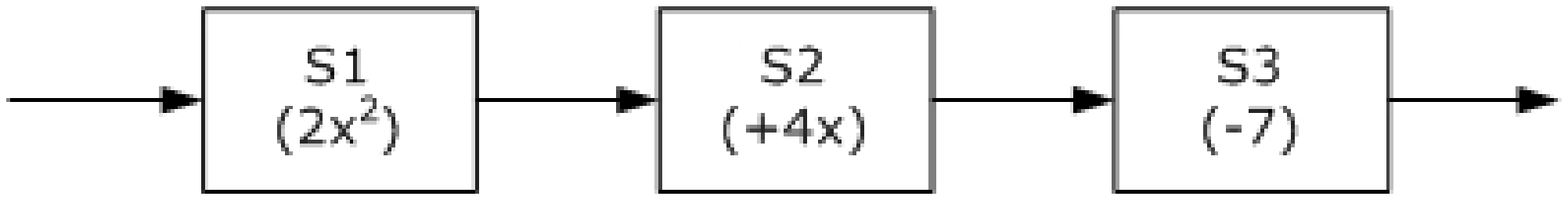}
}
\vspace{1em}

Stage S$_1$ computes $2x^2$ which feeds S$_2$ adding
$4x$, which finally feeds S$_3$ to subtract 7.
A register-transfer level (RTL) implementation requires multiplexors, latches, and clock
inputs on each resource considerably cluttering up the design and model.
%\vspace{1em}
%\mbox{
%  \includegraphics[scale=0.4]{simple_elabed}
%}
%\vspace{1em}
Requiring the user to model these artifacts is not helpful and hinders design exploration.
In our library one simply declares the stages, the function each stage computes, and the route
a {\em transaction} follows through the pipeline.
\begin{verbatim}
    Resource<F1> s1; Resource<F2> s2; Resource<F3> s3;
    Pipeline p = s1 >> s2 >> s3;
\end{verbatim}
Pipeline stages are declared to be resources that are parameterized on a small class that
implements the computational aspect of the stage. The class {\tt Resource} is a
{\em proxy class} for a highly configurable {\tt Stage} class developed in
section \ref{sec:stage}.
The expression above specifies that a
transaction enters the pipeline at {\tt S$_1$}, proceeds to {\tt S$_2$}
then exits the pipeline after {\tt S$_3$}. A modeler can quickly explore a new pipeline where
a stage is repeated (feedback), replicated, or skipped (bypass). For example in a floating-point
multiplication pipeline the adder might be reused consecutively ten times.
In our language this is specified as {\tt Adder*10}.

A key component of our modeling framework are techniques for separating orthogonal
behaviors of the pipeline
into {\em policies} \cite{Alexandrescu01}. The DSL allows us to give a concise configuration of
the pipeline, automatically generate mundane boilerplate code used to
connect modules, insert channels, and generate pipeline control code.
We handle pipelines with arbitrarily complex routing
including feedback and feed-forward (bypass) paths, multi-function, and static or
dynamic pipelines. This generality arises because the DSL is embedded in the host
modeling language (SystemC/C++). This also eliminates the need to write separate language specific
processing phases ({\em e.g.,} lexical analysis, parsing).

We don't claim that the code described here can be used unmodified to model every kind
of pipeline imaginable; that's one of the main reasons we've chosen to embed this in a
general purpose modeling language.  We're motivated by the way a
software design pattern \cite{gof} describes a particular problem that appears over and over
again along with example code
of how to solve the problem (rather than code that works in every context). What we do claim is
that the techniques we describe solve problems that repeatedly appear in pipeline modeling
and that the example code can be reused and modified to suit a particular modeler's needs. Moreover,
the pipeline specifications are compact and efficient allowing a designer to quickly
explore design alternatives.

\subsection{Related Work}
Excellent overviews of pipelining, hardware implementation techniques, and taxonomies are described
in \cite{kogge81,hwang84}.
The compiler research community has developed high-level notations for pipelines to
generate instruction schedulers \cite{bradlee91,Proebsting94}. Our notation
is inspired by that of \cite{Proebsting94}. The work
in \cite{AagaardL93,higgins05,hoffman02} describes notations for specifying pipelines
for downstream tools. Mishra and Dutt \cite{MishraDutt} describe how to validate a pipeline
specification written in the architectural description language \cite{expression}.
Petri Nets \cite {reshadiDuttDate05} and Process Algebra \cite{HarcourtDATE,HarcourtSAS}
have been used to model and simulate pipelines.

We view our research as building on this work in two fundamentally different ways. The first is how
we separate pipeline behaviors into orthogonal policy classes, the second is how these policies are
then configured into a working pipeline with a DSL that is itself embedded in the general purpose
system simulation language SystemC. As \cite{veldhuizen99} points out
external DSLs not embedded in a general purpose language ``tend to have short life-spans due to
limited support and portability, suffer from a lack of tools (particularly debuggers), and it is
usually impossible to use two DSLs in the same source file.''

\subsection{SystemC: Very Briefly}
SystemC is a discrete event modeling and simulation language for designing hardware/software systems
\cite{systemc:lrm,systemc:home}.
SystemC {\em modules} have {\em ports} connected through {\em channels}. SystemC has predefined
channels for hardware like wires ({\tt sc\_signal}) and higher-level channels  such blocking FIFOs.
Users can also define their own channel types. A SystemC module is class
that inherits {\tt sc\_module}. Modules can contain threads ({\tt SC\_THREAD}) or methods
that fire on event changes ({\tt SC\_METHOD}).
SystemC also has a large library of hardware data types including bit vectors and fixed-point types.

\section{System Level Pipeline Specification}
Our pipeline specification framework consists of a small DSL to specify pipeline structure,
and generic models of {\em transactions}, {\em stages}, and {\em transaction routers}. These
components are configured by the user with compile time parameters.
These techniques are inspired by software engineering research in meta-programming \cite{Abrahams05},
generative and generic programming \cite{Czarnecki,GPCE07}, design patterns\cite{gof},
and some advanced C++ programming techniques
\cite{Alexandrescu01,veldhuizen95,vandevoorde03}.

\subsection{Pipeline Specification DSL}
A pipeline expression
defines the route a transaction follows through a pipeline. In a static pipeline this route is fixed.
In a multi-function static pipeline there may be two or more different transaction types each
with a different route. The language provides three binary operators, $>>, +, *$, defined on
pipeline resources. Figure \ref{fig:grammar} shows a small grammar for our DSL.

\begin{figure}[h]
\begin{eqnarray}
\mbox{\textsf{Pipe}} & ::= &  \quid\mbox{\textsf{Term}} \quid\mbox{\texttt{>>}} \quid\mbox{\textsf{Pipe}} \quid | \quid\mbox{\textsf{Term}}\\
\mbox{\textsf{Term}} & ::= &  \quid\mbox{\textsf{Stage}}\quid |
                      \quid\mbox{\textsf{Stage}}\quid * \quid\mbox{\texttt{\textbf{int}}}\quid |
                      \quid\mbox{\textsf{Stage}}\quid + \quid\mbox{\textsf{Stage}} \\
\mbox{\textsf{Stage}} & ::= &  \quid\mbox{\texttt{\textbf{id}}}
\end{eqnarray}
%\hrulefill
%\vspace{-1em}
\caption{\label{fig:grammar} The pipeline DSL grammar}
\end{figure}

Before a stage name can appear in an expression it must be declared as a
{\tt Resource<F>} where {\tt F} is a user defined {\em functor
class}. A functor class encapsulates a function that takes a transaction as an argument and
returns a transaction.

In a pipeline expression
{\tt S$_1$ >> S$_2$ $\ldots$ >> S$_n$},
we call {\tt S$_1$} the pipeline {\em entry} and {\tt S$_n$} the {\em exit}.
The expression {\tt S$_1$ >> S$_2$} indicates
that the output of stage {\tt S$_1$} is fed into stage {\tt S$_2$}.
The expression {\tt S$_1$ * n} indicates that
a stage should be repeated {\tt n} times. This operator represents feedback, not replication of a stage.
The {\tt *} operator is shorthand for repeated
sequencing; {\tt S$_1$ * 3} is shorthand for {\tt S$_1$ >> S$_1$ >> S$_1$}.  Reusing a stage name in an
expression means the stage is also reused and that a transaction is fed back to the stage. For example,
the expression {\tt S$_1$ >> S$_2$ >> S$_3$ >> S$_1$} means that after a transaction exits {\tt S$_3$}
it goes back
to be operated on by {\tt S$_1$} and then exits the pipeline. The expression
{\tt S$_1$ + S$_2$} means that two stages are used in the same cycle and that a
transaction is sent to both
stages. The expression {\tt S$_1$ >> S$_2$ + S$_3$ >> S$_4$} means {\t S$_1$} sends
the transaction to both {\tt S$_2$}
and {\tt S$_3$}, then {\tt S$_2$} and {\tt S$_3$} each then send their transaction to {\tt S$_4$}.
{\tt S$_4$} needs
to know how to handle receiving two transactions simultaneously.

Resource declarations and pipeline expressions are valid C++ and not an external language;
reminiscent of expression templates \cite{veldhuizen95}. To parse
these expressions we overload the {\tt >>}, {\tt +}, and {\tt *} operators and build
an abstract syntax tree which we process to generate SystemC code for
connectivity and control. Pipeline expressions are really compact representations
of {\em reservation tables} \cite{kogge81}.

\begin{figure}
  \includegraphics[scale=0.65]{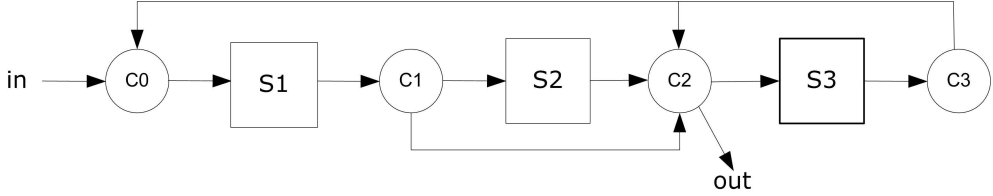}
\caption{\label{fig:complex} Module hierarchy generated by the pipeline
         expression equation \ref{eq:pipeexpr}.}
\end{figure}

Consider the pipeline expression below.
\begin{equation}
  \label{eq:pipeexpr}
  \mbox{{\tt S1 >> S2 >> S3 >> S1 >> S3*2 >> S1 >> S2}}
\end{equation}
Our library generates the SystemC module hierarchy depicted in figure \ref{fig:complex}.
The circles are {\em transaction routers} that are automatically inserted into the module hierarchy.

\section{\label{sec:stage}Abstracting Stages and Transactions}
In addition to the DSL, generic representations of pipeline stages, transactions,
and transaction routers are key
components of the framework. A pipeline is composed of one or more interconnected {\em stages}
that communicate {\em transactions}.
A stage consumes a transaction, operates on it, and sends it on to a subsequent stage
through a router.
A transaction contains user specified {\em data} the stage operates on and {\em control} information
derived from the pipeline expression.
A transaction keeps track of where it is in the pipeline. A transaction router
examines where the transaction currently is, where it has to go, and
uses a lookup table to forward it through the proper port.
Stages have exactly one input and one output, though they may carry complex types.
Routers are multi-ported and handle multiple inputs and outputs of a stage.

Rather than using low-level digital or RTL
modeling constructs ({\em e.g.,} {\tt SC\_METHODS} and {\tt sc\_signals})
we use a {\em transaction level model} (TLM) \cite{glms:systemc}. Pipeline resources are
thread processes that communicate arbitrarily complex transactions through channels
(single place FIFOs) much like
a dataflow simulation \cite{glms:systemc}. We use this framework only for explication,
our classes are not wedded to using threads and FIFOs but are
parameterized on precisely these design choices. By switching policies FIFOs can be
replaced with other SystemC channels
such as Verilog/VHDL like signals (SystemC's {\tt sc\_signal} type) and thread processes with
method processes --- useful in {\em communication refinement} as a modeler
migrates their design to an implementation.

We'll begin developing generic classes using our simple pipeline from the
introduction (section \ref{sec:intro}) as an example. We'll first develop naive implementations of
transaction and stage classes and use these as a basis for our policy based classes.
A pipeline stage needs the original value of
$x$ and the output from the previous stage --- information we'll keep in a transaction.
A transaction also keeps track of its current location in the pipeline.
Figure \ref{fig:trans1} shows an initial version of a transaction. Lines 4-5 specify the data and
lines 7-8 specify control information. The data member {\tt route}
represents the path a transaction follows through the pipeline and is static
because all transactions in a uni-function pipeline share the same route. For a multi-function
pipeline we would have different transaction classes for each pipeline function.
The member {\tt curr} is not static as it represents where a
particular transaction is within the pipeline.
\begin{figure}
\begin{listing}{1}
class Transaction {
public:
   void advance() { curr++; }
   const double orig;
   double data;
private:
   static Route route;
   Route::iterator curr;
};
\end{listing}
%\vspace{-1em}
%\hrulefill
%\vspace{-1em}
\caption{\label{fig:trans1} Naive implementation of a transaction for pipeline in section \ref{sec:intro}.}
\end{figure}

The first stage (naive version) of the pipeline computes $2x^2$ (figure \ref{fig:stage1}).
Lines 3-4 show the stage's port interface, line 11 the function.
Line 12 models a one cycle delay, line 13 advances the transaction to the next stage,
and line 14 writes the modified transaction to the output.
This is all well and good but we'd like to
be able to abstract a stage so that it is as reusable as possible. {\tt Stage}
bundles many design choices into one class and doesn't give a modeler flexibility over
the large number of possible design choices such as functionality, timing, and interface. We'd like
a modeler to be able to {\em configure} a stage to suit a variety of situations.

\begin{figure}
\begin{listing}{1}
class Stage : public sc_module {
public:
  sc_fifo_in<Transaction *> in;
  sc_fifo_out<Transaction *> out;

  Stage() { SC_THREAD(process); }

  void process() {
    while (1) {
      Transaction * t = in.read();
      t->data = t->data + 2 * sqr(t->orig);
      wait(1, SC_NS);
      t->advance();
      out.write(t);
    }
  }
};
\end{listing}
%\vspace{-1em}
%\hrulefill
%\vspace{-1em}
\caption{\label{fig:stage1} Naive implementation of Stage 1 for pipeline in section \ref{sec:intro}.}
\end{figure}

Enter policies, patterns, and generic programming \cite{Alexandrescu01,Czarnecki,Abrahams05}.
``Policies represent configurable behavior for generic functions and types'' \cite{vandevoorde03}.
In C++ a policy is an orthogonal unit of behavior passed as a template argument.
The combination of templates and
multiple inheritance gives us the mechanics to cope with combinatorial behaviors by factoring
out design choices into classes. The main criticism of our {\tt Transaction} and {\tt Stage}
classes are that they
hard-code design choices making them difficult to reuse.

\subsection{Transactions}
The transaction class hard-codes the two data members {\tt orig} and {\tt data} that
are particular to the pipeline; an easy fix with a template parameter (figure \ref{fig:trans2}).
\begin{figure}
\begin{listing}{1}
template <typename T>
class Transaction {
public:
  void advance() { curr++; }
  T value;
  // ... routing code stays same
};
\end{listing}
%\vspace{-1em}
%\hrulefill
%\vspace{-1em}
\caption{\label{fig:trans2} Abstract implementation of a transaction.}
\end{figure}
For our example pipeline instantiating the template parameter {\tt T} with an STL
{\tt pair<double,double>} gets us back the
original transaction with two data members. A {\tt typedef} aids readability.
\begin{verbatim}
typedef Transaction<pair<double, double> > MyTransaction;
\end{verbatim}
%The STL pair type
%can encode an arbitrary number of values; we can have pairs of pairs, and so on
%\cite{vandevoorde03}.
During communication refinement we can instantiate {\tt T} with a SystemC hardware data type.

\subsection{Stages}
{\tt Stage} also hard codes design choices. FIFOs are the
communication model, $2x^2$ is the function it computes, and it all takes one nanosecond.
What if we want our pipeline to be untimed? Substituting zero in {\tt wait} wont do as
an untimed model is different than a zero time model. The non-terminating
while loop implies we're using an {\tt SC\_THREAD} process as opposed to an {\tt SC\_METHOD} process.
All of these design choices can be turned into policies and passed to the {\tt Stage} class as template
parameters. One might wonder
what remains of {\tt Stage}, our host class? ``At an extreme, a host class is totally
depleted of any intrinsic policy. It delegates all design choices and constraints to policies. Such a
host class is a shell over a collection of policies and deals only with assembling the
policies into a coherent behavior'' \cite{Alexandrescu01}.

Decomposing a stage into policies for timing, function, communication, and process yields a
highly configurable class.
Importantly a modeler can implement their own custom policies and does not have to use ours.

\subsubsection{Function Policy}
Creating a policy class for function is a straightforward
application of a functor class\footnote{For readability we name the function {\tt f} as opposed to
overloading the function call operator {\tt ()()}}.
\begin{listing}{1}
template<typename T>
struct TwoSqr {
  static inline T f(T p) {
    p.second = p.second + 2*sqr(p.first);
    return p;
  }
};
\end{listing}
{\tt TwoSqr} declares a function {\tt f} and is parameterized on the type of data in a transaction.
Below is a modified {\tt Stage} that uses a function policy.
Function policies for other stages are analogous.
\begin{listing}{1}
template <class Function>
class Stage : public sc_module,
              public Function {
public:
  // ... as before
    t->data = f(t->data);
  // ... as before
};
\end{listing}
{\em Parameterized inheritance} (line 3) allows us to
call {\tt f} (line 6) in the function policy.

\subsubsection{Communication Policy}
{\tt Stage} hard-codes FIFOs as the communication medium.
Factoring out {\tt Stage}'s port interface into a separate
class is more involved. SystemC's port classes are template classes.
C++ allows us to specify a template class as a template parameter;
 {\em template template parameters} appear frequently in policy based
design \cite{Alexandrescu01}.
\begin{listing}{1}
template<
   typename Transaction,
   template <typename> class InPort,
   template <typename> class OutPort
>
struct StageInterface {
  InPort<Transaction *> in;
  OutPort<Transaction *> out;
};
\end{listing}
{\tt StageInterface} is now parameterized on the transaction (line 2) and the input and output
port interfaces (lines 3-4). A {\tt typedef} helps with readability and gets us back our example stage
interface that uses FIFOs.
\begin{verbatim}
typedef
  StageInterface<MyTransaction, sc_fifo_in, sc_fifo_out> FIFOInterface;
\end{verbatim}

\subsubsection{Timing Policy}
{\tt Stage} hard-codes a performance model of a one nanosecond delay.
A trivial way
to generalize this is to allow the user to pass an integer through the constructor and use that as
the delay. This assumes that the performance model will be a simple wait statement and nothing more
complicated; not very general. Additionally we might want to support an untimed model where we
would expect there to be no simulation performance penalty in calling {\tt wait}.
One way to do that is to ensure that the call to {\tt wait} is removed by the
compiler for untimed models.
We define two timing policies {\tt TimedPolicy} and
{\tt UntimedPolicy} fully aware that the modeler could design more complicated policies.
\begin{verbatim}
struct TimedPolicy {
  inline static void wait(int t) { ::wait(t, SC_NS); }
};

struct UntimedPolicy {
  inline static void wait(int t) { }
};
\end{verbatim}
If {\tt UntimedPolicy} policy is instantiated
the call to {\tt wait} is inlined with an empty function body, eliminating
function call overhead.

\subsubsection{Our Host {\tt Stage} Class}
Having defined several orthogonal policies figure
\ref{fig:stage} shows our a generic stage class.
\begin{figure}
\centering
\begin{listing}{1}
template <class Transaction,
          class Function,
          class DelayModel,
          class PortInterface>
class Stage : public sc_module,
              public Function,
              public DelayModel,
              public PortInterface {
public:
  Stage() { SC_THREAD(process); }

  void process() {
    while(1) {
      Transaction * t = in.read();
      t->data = f(t->data);
      wait(1);
      t->advance();
      out.write(t);
    }
  }
};
\end{listing}
%\hrulefill
%\vspace{-1em}
\caption{\label{fig:stage}Our generic policy based pipeline stage class.}
\end{figure}
This new stage class goes a long way in being generic and reusable. However the non-terminating
{\tt while} loop in the {\tt process} function is not generic; it implies we're using
thread processes ({\tt SC\_THREAD}) as opposed to method processes ({\tt SC\_METHOD}), or
clocked threads ({\tt SC\_CTHREAD}). While we don't have space to show it here we also factor
out the {\em process policy} into {\tt ThreadPolicy} and {\tt MethodPolicy}
adding one more template parameter {\tt ProcessPolicy} to the {\tt Stage} class.

\subsection{Putting it all together}
Recall that our DSL initially uses a proxy class {\tt Resource} for stages.
To generate a complete simulation model we pass our policy classes to {\tt Resource}.
\begin{verbatim}
Resource<Sqr<Data>, MyTransaction, TimedPolicy, FIFOInterface, Threading> s1;
\end{verbatim}
To shorten this up a bit we can give reasonable default values to the template parameters or use
an extra configuration repository class \cite{Czarnecki}.

We don't have room to show the {\em transaction routers}. These are multi-ported
modules parameterized on a transaction and use a vector indexed by the stage number
to lookup the appropriate channel to write the transaction to. We have not yet decomposed these
further into other policies. Pipelines that demand a more complicated resource arbitration
scheme require an {\em arbitration policy}.

\section{Conclusions and Future Work}
We've presented techniques for modeling and simulating system level pipelines. A small DSL
gives a compact representation of the pipeline and enables us to automatically
generate tedious boilerplate and control code. Generic representations of transactions and stages
decomposed into policy classes allow us to reuse large amounts of code used to describe pipeline
resources.

One aspect we haven't addressed is {\em when} a transaction can be initiated in the pipeline, the
{\em issue latency}. Issue latencies are derivable from the DSL; \cite{kogge81} shows us how and
\cite{Proebsting94} makes it fast. One area of future work are policies for gathering
performance statistics as well as policies for generating implementation level models. More
policies will be discovered as we model more complicated pipelines, including pipelines
that use global state, such as processor instruction pipelines with instruction and data caches.
In terms of abstractions used in our framework {\em concept models}
\cite{concepts} can help clarify requirements of our policies.

The pipeline DSL needs enhancing. For example, a stage replication
operator {\tt S$_1\hspace{0.1mm}^\wedge$3} could mean {\em
replicate} hardware; instantiate {\tt S$_1$} three times as
opposed to feedback ({\tt S$_1$*3}). At the moment parentheses are
meaningless but giving semantics to expressions such as {\tt
((S$_1$ >> S$_2$)*2 >> S$_3$)*3} by ``unrolling'' makes sense.
Also, we could probably make pipeline descriptions even more
concise by using the Boost lambda library \cite{boostlambda} for
specifying functors.

\bibliographystyle{plain}
\bibliography{arxiv}

\end{document}